\documentclass[copyright,creativecommons]{eptcs}
 \providecommand{\event}{CCA 2010} % Name of the event you are submitting to
 \usepackage{breakurl}             % Not needed if you use pdflatex only.
\usepackage{amsmath}
\usepackage{amssymb}
\usepackage{amsthm}

\newtheorem{example}{Example}

\title{Computation with Advice}
\author{
 Vasco Brattka
\institute{Laboratory of Foundational Aspects of Computer Science\\
             Department of Mathematics \& Applied Mathematics\\
             University of Cape Town, South Africa\\
}
\email{Vasco.Brattka@uct.ac.za}
\and
Arno Pauly\thanks{Corresponding Author}
\institute{Computer Laboratory\\ University of Cambridge, United Kingdom}
\email{Arno.Pauly@cl.cam.ac.uk}
}
\def\titlerunning{Computation with Advice}
\def\authorrunning{V. Brattka \& A. Pauly}

\begin{document}
\theoremstyle{definition}
\newtheorem{theorem}{Theorem}
\newtheorem{definition}[theorem]{Definition}
\newtheorem{lemma}[theorem]{Lemma}
\newtheorem{problem}[theorem]{Problem}
\newtheorem{proposition}[theorem]{Proposition}
\newtheorem{corollary}[theorem]{Corollary}
\newtheorem{deflemma}[theorem]{Definition \& Lemma}
\newcommand{\dom}{\operatorname{dom}}
\newcommand{\Lev}{\operatorname{Lev}}
\newcommand{\range}{\operatorname{range}}
\newcommand{\id}{\textnormal{id}}
\newcommand{\Baire}{\mathbb{N}^\mathbb{N}}
\newcommand{\Cantor}{\{0, 1\}^\mathbb{N}}
\newcommand{\mto}{\rightrightarrows}
\newcommand{\Sep}{\textnormal{Sep}}

\maketitle

\begin{abstract}
Computation with advice is suggested as generalization of both computation with discrete advice and Type-2 Nondeterminism. Several embodiments of the generic concept are discussed, and the close connection to Weihrauch reducibility is pointed out. As a novel concept, computability with random advice is studied; which corresponds to correct solutions being guessable with positive probability. In the framework of computation with advice, it is possible to define computational complexity for certain concepts of hypercomputation. Finally, some examples are given which illuminate the interplay of uniform and non-uniform techniques in order to investigate both computability with advice and the Weihrauch lattice.
\end{abstract}

\section{Introduction}
An approach to classify the incomputability of some problem is what kind of help we need to overcome it and compute the problem. The general idea is that for some (generally incomputable) function $f$, we find a computable function $g$, such that $f(x) = g(x, w)$ holds for suitable $w$. One can conceive many potential requirements for the \emph{suitability} of $w$. Concepts discussed so far in the literature are computability with discrete advice, where basically the set of potential advice values $w$ is restricted (\cite{ziegler5}), non-determinism, which amounts to wrong advice being effectively recognizable (\cite{ziegler2}, \cite{paulybrattka}), and the uniqueness of suitable advice. Others will be newly introduced in the following.

For computability with discrete advice, preservation under Weihrauch reducibility was shown directly (\cite{paulyreducibilitylattice}), for the other concepts it was concluded from various arguments, however, as we will see, there is a uniform proof in the generic case. This is particularly useful, as we can extend a small number of negative results via Weihrauch reducibility to a huge number of examples.

On the other hand, establishing computability with a certain type of advice for a given problem directly implies various results of non-reducibility in the Weihrauch lattice. Thus, the interplay of computability with advice and Weihrauch reducibility allows to replace negative proofs -- which tend to be hard -- by positive ones --which are straight-forward in many cases-- in all but a few cases.

Finally, understanding concepts such as non-determinism as special cases of computability with advice opens up a way of introducing computational complexity for them by considering the least computational complexity of some $g$ witnessing the computability with advice of $f$ as the complexity of $f$ regarding this type of advice. We will demonstrate how this allows us to pose and answer the $\textbf{P} = \textbf{NP}?$ problem for Type-2 Machines.

\section{Definitions}
\label{secdefinitions}
There are at least two distinct approaches we could use to define computation with advice. Considering an arbitrary (multi-valued) function $f : \subseteq X \mto Y$ between represented spaces, and an additional represented space $Z$ serving as advice space, we can use computable functions $g : \subseteq X \times Z \mto Y$ as potential witnesses of computability with advice of $f$. Alternatively, we can introduce computability with advice as a generalization of realizers. We will pick the latter approach here, the reasons for and consequences of that shall be discussed in Section \ref{secrealizers}.

For a (multi-valued) function $f: \subseteq (X, \delta_X) \mto (Y, \delta_Y)$ between represented spaces, we call $F : \subseteq \Baire \to \Baire$ a realizer of $f$ (denoted by $F \vdash f$), if $\delta_Y \circ F(x) \in f \circ \delta_X(x)$ holds for all $x \in \dom(f\delta_X)$. In the following, we will usually not list the representations, but simply assume that $X$ is represented by $\delta_X$, and so on. Given a set $X$, $\mathcal{P}(X)$ shall denote its power set.

\begin{definition}
\label{defgeneric}
Let $Z$ be a represented space, and $\mathcal{A} \subseteq \mathcal{P}(Z)$ a set of subsets of $Z$ with $\emptyset \notin \mathcal{A}$. We call a function $f : \subseteq X \mto Y$ computable with $(Z, \mathcal{A})$-advice, if there is a computable function $F : \subseteq \Baire \times \Baire \to \Baire$ and a family $(A_x)_{x \in \dom(f\delta_X)}$ in $\mathcal{A}$ such that $\delta_Y(F(x, y)) \in f(\delta_X(x))$ holds for $\delta_Z(y) \in A_x$.
\end{definition}

We will introduce some shortcuts for useful choices of $\mathcal{A}$ in the following. The list below is not meant to be exhaustive, many other systems $\mathcal{A}$ are conceivable to give rise to interesting concepts. Some of the sets $\mathcal{A}$ we consider will produce trivial notions of computability with $(Z, \mathcal{A})$-advice, but produce interesting concepts once combined with effective advice, which shall be introduce later.

\begin{definition}
\label{defgeneric2}
A function $f : \subseteq X \mto Y$ is called computable with $Z$-advice, if $f$ is computable with $(Z, \mathcal{P}(Z) \setminus \{\emptyset\})$-advice.
\end{definition}
The Definition \ref{defgeneric2} is very similar to \cite[Definition 48]{ziegler5}. However, due to the use of realizers instead of a direct approach, our notion of computability with $Z$-advice corresponds to what might be called computability with \emph{weak} $Z$-advice in the terms of \cite{ziegler5}. For represented spaces admitting an injective representation, both notions are identical. In addition, for many natural examples of functions $f$, $f$ tends to be computable with $Z$-advice in either both senses or in neither. A counterexamples is given as \cite[Example 23]{ziegler5} (reproduced here in Section \ref{secrealizers}).

Not all spaces give rise to an interesting notion of computability with $Z$-advice. For example, every function is computable with $\Cantor$-advice, with $\mathbb{R}$-advice and with $\Baire$-advice, as all these spaces admit a computable partial surjection onto Baire space. As demonstrated in \cite{ziegler5}, each finite discrete space, the discrete space $\mathbb{N}$ as well as the Sierpi{\'n}ski-space all induce distinct concepts of computability with advice.

Strengthening \cite[Proposition 8e]{ziegler5}, stating that computability with $\{1, \ldots, k\}$-advice implies non-uniformly computability, we present the following proposition, where $\leq_T$ shall denote Turing reducibility.
\begin{proposition}
\label{nonuniform}
A function $f : \subseteq X \mto Y$ is computable with $\mathbb{N}$-advice, if and only if for each $x \in \dom(f\delta_X)$ there is an $y \in \delta_Y^{-1}(f(\delta_X(x)))$ with $y \leq_T x$.
\begin{proof}
Assume $f$ is computable with $\mathbb{N}$-advice, let $\delta_\mathbb{N}$ be an injective standard representation of $\mathbb{N}$, and let $F$ be the witness according to Definition \ref{defgeneric}. Then pick some $n \in A_x$, and consider $y := F(x, \delta_\mathbb{N}^{-1}(n))$, this fulfills $y \leq_T x$. On the other hand, for each $x$ let $n_x$ be the index of a Turing functional witnessing $y \leq_T x$ for a suitable $y$. Then the sets $(\{n_x\})_{n \in \mathbb{N}}$ together with a universal machine as $F$ witness computability of $f$ with $\mathbb{N}$-advice.
\end{proof}
\end{proposition}

On a represented space $(X, \delta)$ a topology is defined canonically as the final topology induced by the standard topology on Baire space along $\delta$. The set of open sets can be represented by setting $\delta^\mathcal{O}(p) = U$, if $p$ encodes a list of all finite words $w$ with $w\Baire \cap \dom(\delta) \neq \emptyset$ and $\delta(w\Baire \cap \dom(\delta)) \subseteq U$. From that, a representation for closed sets is derived by $\delta^\mathcal{C}(p) = X \setminus \delta^\mathcal{O}(p)$.
\begin{definition}
\label{defclopen}
A function $f : \subseteq X \mto Y$ is called computable with closed (open, clopen) $Z$-advice, if $f$ is computable with $(Z, \mathcal{C}(Z))$-advice (with $(Z, \mathcal{O}(Z))$-advice, with $(Z, \mathcal{C}(Z) \cap \mathcal{O}(Z))$-advice), where $\mathcal{C}(Z)$ $(\mathcal{O}(Z))$ denotes the set of non-empty closed (open) subsets of $Z$.
\end{definition}

If $Z$ is a $T_1$-space, closed $Z$-advice is equivalent to $Z$-advice: It follows from the Axiom of Choice that for every family of non-empty subsets $A_x$ of $Z$ there is a family of points $z_x\in A_x$ and if $Z$ is a $T_1$-space, then the singletons $\{z_x\}$ are closed.
 For the Sierpi{\'n}ski space $\mathbb{S}$, however, computability with closed $\mathbb{S}$-advice already implies computability, as there is a unique minimal closed non-empty subset in each of these spaces, which contains a computable element.

Open $Z$-advice is equivalent to $Z$-advice for discrete spaces. The existence of a unique minimal open set such as in the Sierpi{\'n}ski space leads to an equivalence of computability with open $Z$-advice and computability. Many other natural spaces are covered by the following result:
\begin{proposition}
Let $Z$ be an infinite computable metric spaces. Then open $Z$-advice is equivalent to $\mathbb{N}$-advice.
\begin{proof}
If $Z$ is a computable metric space, there is a countable dense subset which can be considered as the range of some computable partial function $\nu : \subseteq \mathbb{N} \to Z$. As every non-empty open subset $U$ of $Z$ contains some $\nu(n_U)$, the number $n_U$ can be used at advice rather than some element of $U$.

For the other direction, observe that an infinite computable metric space admits a countable family $(U_n)_{n \in \mathbb{N}}$ of disjoint non-empty open sets. In the subspace given by the union of these sets, each individual open set is also closed. This turns the function $\chi : \subseteq Z \to \mathbb{N}$ with $\chi(x) = n$ for each $x \in U_n$ computable. Thus, the number $n$ as advice can be replaced by the set $U_n$.
\end{proof}
\end{proposition}
This remains true for clopen $\Cantor$-advice and clopen $\Baire$-advice. Due to lack of non-trivial clopen subsets, computability with clopen $\mathbb{R}$-advice already implies computability.

\begin{definition}
A function $f : \subseteq X \mto Y$ is called computable with unique $Z$-advice, if $f$ is computable with $(Z, \mathcal{U}(Z))$-advice, where $\mathcal{U}(Z) := \{ \{z\} \mid z \in Z\}$ denotes the set of singletons in $Z$.
\end{definition}

As every non-empty set contains a singleton, unique $Z$-advice is equivalent to $Z$-advice.

For the next definition, our space $Z$ has to be equipped with a measure. Hence, effective topological measure spaces (e.g. \cite{hertling2}) are a natural setting. The standard examples are presented and studied in Section \ref{secrandom}.

\begin{definition}
A function $f : \subseteq X \mto Y$ is called computable with random $Z$-advice, if $f$ is computable with $(Z, \mathcal{R}(Z))$-advice, where $\mathcal{R}(Z)$ denotes the set of subsets with positive measure.
\end{definition}

This approach is somewhat complementary to unique $Z$-advice: Instead of smallness as decisive criterion, now the sets have to be large. The intuition behind our definition is that by randomly guessing a name $w$ of some $\delta_Z(w) \in Z$, and then computing $F(x, w)$, we have a positive probability of arriving at the intended value $f(\delta_X(x))$. From this point of view, requiring a certain minimal measure $\varepsilon > 0$ rather than just positive measure seems to be a good choice, however, this notion is (presumably) less stable.

If the set $\mathcal{A}$ can be represented in any natural way, we can introduce effective advice. Typical examples here are closed or open subsets of a represented space, with the representations introduced before Definition \ref{defclopen}, or subclasses thereof represented by restrictions of these representations. The following definition generalizes \cite[Definition 7.1]{paulybrattka}, which encompasses the special case of $Z$ being a subspace of $\Baire$, and $\mathcal{A}$ being $\mathcal{C}(Z)$.

\begin{definition}
\label{defeffective}
Let $Z$ be a represented space, and $\mathcal{A} \subseteq \mathcal{P}(Z)$ a represented set of subsets of $Z$ with $\emptyset \notin \mathcal{A}$. We call a function $f : \subseteq X \mto Y$ computable with effective $(Z, \mathcal{A})$-advice, if there are computable functions $F : \subseteq \Baire \times \Baire \to \Baire$ and $A : \subseteq \Baire \to \mathcal{A}$ such that $\delta_Y(F(x, y)) \in f(\delta_X(x))$ holds for $\delta_Z(y) \in A(x)$.
\end{definition}

If $f$ is computable with effective $(Z, \mathcal{A})$-advice, we can use the computable function $A$ to compute $A(x)$, given some $x \in \dom(f\delta_X)$. If we now could pick a name $y$ of some element $\delta_Z(y) \in A(x)$, an application of the computable function $F$ would yield a name of $f(\delta_X(x))$. Thus, the only problem here is to choose an element from a set from $\mathcal{A}$.

Choosing from an open set is always computable: Each finite sequence $w$ listed in a $\delta^\mathcal{O}$-name of $U$ is a prefix of some $\delta$-name of an element $x \in U$, and provided that $U$ is non-empty, for any valid prefix, a strictly longer one is also listed. Thus computability with effective open $Z$-advice directly implies computability, hence, this notion is of no further interest.

For closed sets, however, the situation is different, given that our representation provides only negative information. As demonstrated in \cite{paulybrattka}, this concepts leads to very interesting behaviour. In the remainder of the paper, effective advice is considered only for systems of sets $\mathcal{A} \subseteq \mathcal{C}(Z)$.

Therefore, studying effective unique $Z$-advice and effective random $Z$-advice amounts to effective $(Z, \mathcal{U}(Z) \cap \mathcal{C}(Z))$-advice and effective $(Z, \mathcal{R}(Z) \cap \mathcal{C}(Z))$-advice. The former makes sense particularly for $T_1$-spaces, where each singleton is closed; and the latter for spaces equipped with a Borel measure, where each closed set is at least measurable.

\section{Connections to Weihrauch Reducibility}
\label{secweihrauch}
From Definition \ref{defgeneric}, the connection to Weihrauch reducibility is rather straight-forward. For the background on Weihrauch reducibility, we refer to \cite{brattka2}, and only present the definition. By $\langle \ \rangle$, we denote the usual pairing operation on Baire space, and by $\id$ the identity function on Baire space.

\begin{definition}
\label{defweihrauchreduc}
For two (multi-valued) functions $f$, $g$ between represented spaces, $f \leq_W g$ holds, if there are computable functions $F$, $K$, such that for each $G \vdash g$ we also have $F\circ \langle \id, GK\rangle \vdash f$.
\end{definition}

\begin{theorem}
\label{theoweihrauch}
Let $f : \subseteq X_1 \mto Y_1$ and $g : \subseteq X_2 \mto Y_2$ be (multi-valued) functions between represented spaces. If $g$ is computable with $(Z, \mathcal{A})$-advice and $f \leq_W g$ holds, then $f$ is also computable with $(Z, \mathcal{A})$-advice.
\begin{proof}
Due to the assumption, there is a $G : \subseteq \Baire \times \Baire \to \Baire$ and a family $(A_x)_{x \in \dom(g\delta_{X_2})}$ in $\mathcal{A}$, such that $\delta_{Y_2}(G(x, y)) \in g(\delta_{X_2}(x))$ holds for $\delta_Z(y) \in A_x$. Now let $f \leq_W g$ be witnessed by computable functions $F$, $K$; and consider the computable function $H : \subseteq \Baire \times \Baire \to \Baire$ defined by $H(x, y) = F\langle x, G(K(x), y)\rangle$.

If $\delta_Z(y) \in A_{K(x)}$ holds, we have $\delta_{Y_2}(G(K(x), y)) \in g(\delta_{X_2}(K(x)))$. By the Axiom of Choice, $g$ has a realizer $G'$ and, in particular,
    one with $G'K(x)=G(K(x),y)$. But then, due to the properties of $F$, also $\delta_{Y_1}(H(x, y)) \in f(\delta_{X_1}(x))$ follows.
\end{proof}
\end{theorem}

Theorem \ref{theoweihrauch} remains true for effective $(Z, \mathcal{A})$-advice. The computable function $A'$ for $f$ is just obtained from the computable function $A$ for $g$ via $A' = A \circ K$.

\begin{theorem}
A function $f : \subseteq X \mto Y$ is computable with $Z$-advice, if and only if there is a function $g : \subseteq \Baire \mto Z$ with $f \leq_W g$. The function $g$ can also be chosen as a single-valued one.
\begin{proof}
First of all, each function $g : \subseteq W \mto Z$ is computable with $Z$-advice: We consider the computable projection $\pi_2 : \Baire \times \Baire \to \Baire$ defined by $\pi_2(x, y) = y$, together with the sets $A_x = \{z \in Z \mid z \in g(\delta_W(x))\}$. Now application of Theorem \ref{theoweihrauch} gives one of the implications.

For the other direction, assume that $f : \subseteq X \mto Y$ is computable with $Z$-advice, witnessed by the computable function $F$ together with the sets $(A_x)_{x \in \dom(f\delta_X)}$. As each set $A_x$ is non-empty, we can invoke the Axiom of Choice to obtain a choice function $g : \subseteq \Baire \to Z$ with $g(x) \in A_x$ for all $x \in \dom(f\delta_X)$. Now we have $F\langle \id, G\rangle \vdash f$ for every $G \vdash g$; thus, we have $f \leq_W g$.
\end{proof}
\end{theorem}

Effective advice was studied in detail in \cite{paulybrattka}. Besides a characterization of effective advice as non-determinism as in \cite{ziegler2}, effective $Z$-advice also admits a deep connection to Weihrauch reducibility. To express it, we introduce the closed choice on some represented space $X$ as the multi-valued function $C_X : \subseteq \mathcal{C}(X) \mto X$ which satisfies $\dom(C_X) = \{A \in \mathcal{C}(X) \mid A \neq \emptyset\}$ and $C_X(A) = A$. For any system $\mathcal{A} \subseteq \mathcal{C}(X)$, let $C_X^\mathcal{A}$ be the restriction of $C_X$ to $\mathcal{A}$.

\begin{theorem}
\label{effectiveadvicechoice}
For $\mathcal{A} \subseteq \mathcal{C}(Z)$, a function $f$ is computable with effective $(Z, \mathcal{A})$-advice, if and only if $f \leq_W C_Z^\mathcal{A}$ holds.
\begin{proof}
The problem $C_Z^\mathcal{A}$ clearly is computable with effective $(Z, \mathcal{A})$-advice. The computable witness $F$ is the projection to the second component, the computable witness $A$ is the relevant restriction of $\delta^\mathcal{C}$, if $\delta$ is the representation of $Z$. Theorem \ref{theoweihrauch} thus provides one direction of the equivalence.

For the other direction, assume that the computability with effective $(Z, \mathcal{A})$-advice of $f$ is witnessed by computable $F$ and $A$, as before. Let $B$ be a computable realizer of $A$. If $C$ is a realizer of $C_Z^\mathcal{A}$, then $F\langle \id, CB\rangle \vdash f$ follows rather directly from the definitions. Thus, $f \leq_W C_Z^\mathcal{A}$ is established.
\end{proof}
\end{theorem}
For $C_X^{\mathcal{U}(X) \cap \mathcal{C}(X)}$ the short form $UC_X$ was introduced in \cite[Section 6]{paulybrattka}. In a similar fashion, we shall use $PC_X$ as abbreviation of $C_X^{\mathcal{R}(X) \cap \mathcal{C}(X)}$.

We conclude this section by recalling some results from \cite{paulybrattka} regarding the positions in the Weihrauch lattice of closed choice and unique closed choice for various spaces. For finite discrete spaces, we have $C_{\{1, \ldots, n\}} \equiv_W MLPO_n$, with $MLPO_n$ originally introduced in \cite{weihrauchc}. In particular, computability with effective $\{1, \ldots, n\}$-advice of $f$ implies $\textnormal{Lev}(f) \leq n$, while the converse is false. Here, $\textnormal{Lev}(f)$ is the Level of a function introduced in \cite{hertling}. Computability with unique effective $Z$-advice for a finite space $Z$ already implies computability.

The functions computable with effective $\Cantor$-advice are exactly the weakly computable functions, computability with unique effective $\Cantor$-advice implies computability. Effective $\mathbb{N}$-advice is incomparable with effective $\Cantor$-advice, and equivalent to unique effective $\mathbb{N}$-advice. The functions computable with effective $\mathbb{N}$-advice are exactly the functions computable with finitely many mind-changes (\cite{ziegler3}, \cite{debrecht}).

Effective $(\mathbb{N} \times \Cantor)$-advice is equivalent to effective $\mathbb{R}$-advice, and obviously is stronger than both effective $\mathbb{N}$-advice and effective $\Cantor$-advice. Unique effective $\mathbb{R}$-advice is equivalent to effective $\mathbb{N}$-advice.

Finally, all effectively Borel measurable functions are computable with effective $\Baire$-advice, and every single-valued function with Polish domain which is computable with effective $\Baire$-advice is Borel measurable.

\section{Changing the Advice Space}
How is $(Z_1, \mathcal{A}_1)$-advice linked to $(Z_2, \mathcal{A}_2)$-advice for different represented spaces $Z_1$ and $Z_2$ and/or different systems of subsets $\mathcal{A}_1$ and $\mathcal{A}_2$? A partial answer is given in the following proposition:

\begin{proposition}
If there is a computable surjection $\iota :\subseteq Z_1 \to Z_2$, such that for all $A \in \mathcal{A}_2$ there is an $A' \in \mathcal{A}_1$ with $A' \subseteq \iota^{-1}(A)$, then computability with $(Z_2, \mathcal{A}_2)$-advice implies computability with $(Z_1, \mathcal{A}_1)$-advice
\begin{proof}
Let $J:\subseteq\mathbb{N}^\mathbb{N}\to\mathbb{N}^\mathbb{N}$ be a
computable realizer of $\iota$ and let
$F_2:\subseteq\mathbb{N}^\mathbb{N}\times\mathbb{N}^\mathbb{N}\to\mathbb{N}^\mathbb{N}$
be a computable
witness for the fact that $f$ is computable with
$(Z_2,\mathcal{A}_2)$-advice.
Let $(A_x)_{x\in\dom(f\delta_X)}$ be a corresponding family of sets in
$\mathcal{A}_2$.
By assumption and the Axiom of Choice there exists a family
$(A'_x)_{x\in\dom(f\delta_X)}$ of sets in $\mathcal{A}_1$ such that
$A'_x\subseteq\iota^{-1}(A_x)$. This family together with the computable
function
$F_1$ defined by $F_1(x,y):=F_2(x,J(y))$ witness the fact that $f$ is
computable with $(Z_1,\mathcal{A}_1)$-advice.
\end{proof}
\end{proposition}

If $\mathcal{A}_1$ and $\mathcal{A}_2$ are of the same type, usually a total computable surjection is sufficient. Note that this proposition also provides interesting results in the case $Z_1 = Z_2$, in particular, it contains a number of implications between different types of advice given earlier as special cases.

In order to extend the result to any kind of effective advice, there must be a computable function $\tau : \mathcal{A}_2 \to \mathcal{A}_1$ with $\tau(A) \subseteq \iota^{-1}(A)$ for all $A \in \mathcal{A}_2$. In the case that both $\mathcal{A}_1$ and $\mathcal{A}_2$ are the system of closed sets, this was shown as \cite[Proposition 3.7]{paulybrattka}.

In order to cover the composition of functions, a composition of advice will be introduced:
\begin{definition}
\label{defcompositionadvice}
For $\mathcal{A} \subseteq \mathcal{P}(Z) \setminus \{\emptyset\}$ and $\mathcal{B} \subseteq \mathcal{P}(Y) \setminus \{\emptyset\}$, define $\mathcal{A} \circ \mathcal{B} \subseteq \mathcal{P}(Z \times Y) \setminus \{\emptyset\}$ via: $$\mathcal{A} \circ \mathcal{B} := \left \{\bigcup_{x \in B} A_x \times \{x\} \mid B \in \mathcal{B} \ \forall x \in B \ A_x \in \mathcal{A} \right\}$$
\end{definition}

\begin{lemma}
\label{lemmacomposition}
Let $f : \subseteq X \mto Y$ be computable with $(Z_1, \mathcal{A})$-advice, and let $g : \subseteq W \mto X$ be computable with $(Z_2, \mathcal{B})$-advice. Let $Z_2$ be equipped with an injective representation. Then $f \circ g$ is computable with $(Z_1 \times Z_2, \mathcal{A} \circ \mathcal{B})$-advice.
\begin{proof}
Let $F$ be the computable witness for $f$, together with the sets $(A_x)_{x \in \dom(f\delta_X)}$; and let $G$ be the computable witness for $g$, together with the sets $(B_w)_{w \in \dom(g\delta_W)}$. Then $\delta_Y(F(G(w, y),z)) \in (f \circ g)(\delta_W(w))$ holds for $\delta_{Z_2}(y) \in B_w$ and $\delta_{Z_1}(z) \in A_{G(w, y)}$. If $\delta_{Z_2}$ is injective, the set $$\bigcup_{v \in B_w} A_{G(w, \delta_{Z_2}^{-1}(v))} \times \{v\} \in \mathcal{A} \circ \mathcal{B}$$ is the set of suitable advice.
\end{proof}
\end{lemma}

A particularly nice formulation of the preceding lemma is obtained in the case $\mathcal{B} = \mathcal{U}(Z_2)$, due to $\mathcal{A} \circ \mathcal{U}(Z_2) = \mathcal{A} \times \mathcal{U}(Z_2)$. This in turn is a generalization of \cite[Lemma 12b]{ziegler5}. In the case of effective advice, the sets $A_x$ depend computably on the set $B$ in Definition \ref{defcompositionadvice}, hence much more can be said. We refer to the Independent Choice Theorem (\cite[Theorem 7.3]{paulybrattka}), which shows that effective closed choice is closed under composition.

Products and coproducts of functions are compatible with products and coproducts of the advice spaces; and this even for generic advice types $\mathcal{A}$, yielding the following lemma:
\begin{lemma}
\label{lemmaproductscoproducts}
Let $f : \subseteq X_1 \mto Y_1$ be computable with $(Z_1, \mathcal{A}_1)$-advice, and let $g : \subseteq X_2 \mto Y_2$ be computable with $(Z_2, \mathcal{A}_2)$-advice.
\begin{enumerate}
\item  Then $f \times g$ is computable with $(Z_1 \times Z_2, \mathcal{A}_1 \times \mathcal{A}_2)$-advice.
\item  Then $f \sqcup g$ is computable with $(Z_1 \sqcup Z_2, \mathcal{A}_1 \sqcup \mathcal{A}_2)$-advice.
\end{enumerate}
\end{lemma}

For the special cases we have considered so far, that is closed, open sets, singletons and sets with positive measure, products or disjoint unions of such sets are of the same type in either the product or the coproduct (except singletons) of the spaces. For spaces that are computably homeomorphic to their product or coproduct, the statement of Lemma \ref{lemmaproductscoproducts} takes an even nicer form. For example, if $f$ and $g$ are computable with random $\Cantor$-advice, then so is $f \times g$.
\section{On Definitions via Realizers}
\label{secrealizers}
While many results do not depend on the question whether the involved notions are defined via realizers or directly, others do. In particular, a consistent choice is necessary:
\begin{example}[{\cite[Example 23]{ziegler5}}]
Let $\mathcal{S}^1$ be the circle obtained from the unit interval $[0, 1]$ by identifying $0$ and $1$ with $\iota$ as quotient map. The standard representation of $\mathcal{S}^1$ is $\iota \circ \rho$ with the standard representation $\rho$ of the unit interval. Define a function $f: \mathcal{S}^1 \to [-1, +1]$ by $f(\iota(x)) = x - 1$ for $x \in [0, 1] \setminus \mathbb{Q}$ and $f(\iota(x)) = x$ for $x \in [0, 1[ \cap \mathbb{Q}$.

$\left \{ \rho^{-1}([0, 1[ \cap \mathbb{Q}), \rho^{-1}(\{1\} \cup ([0, 1] \setminus \mathbb{Q})) \right \}$ is a 2-partition of $\dom(\iota\rho)$, such that $f$ has a realizer $F$ that is computable restricted to each set. Hence, $f$ is computable with $\{1, 2\}$-advice.

However, there is no partition $\{A, B\}$ of $\mathcal{S}^1$ rendering both $f_{|A}$ and $f_{|B}$ computable.
\end{example}
Every function is trivially Weihrauch reducible to each of its realizers, this shows that Theorem \ref{theoweihrauch} would not hold anymore, if we had chosen to work on the spaces of interest directly, rather than using representations as in Definition \ref{defgeneric}.

Of course, rather than using Weihrauch reducibility as defined in Definition \ref{defweihrauchreduc}, both the reduction and the advice could have been defined directly. An analogue to Theorem \ref{theoweihrauch} would have been obtainable in that framework. Direct reducibilities have been considered e.g. in \cite{paulyreducibilitylattice}, and they have a quite similar overall structure to the realizer-version suggested in \cite{brattka2}, \cite{brattka3}. In fact, the realizer-version can be considered as a substructure of the direct reducibilities, by restricting these to multi-valued functions on Baire space. This circumstance also allows to apply many of the results in \cite{paulyreducibilitylattice} to Weihrauch reducibility as understood here.

In general, whenever the interest lies in computational aspects, the realizer-versions of definitions are preferable. As all actual computations have to be executed on names rather than on objects, we should exploit this circumstance, and e.g. treat different names of the same object differently, if this allows us to compute things otherwise unobtainable. If the purely topological notions are the main focus, the direct definitions might be more appropriate.

Regarding the work presented here, there is another reason in favour of the realizer-definition of computation with advice. All negative results for the realizer-version also apply to the direct definition. By this circumstance, Weihrauch reducibility is suitable to prove lower bounds for the cardinality of discontinuity in \cite{ziegler5}.

A (small) drawback for the realizer-definition can be found in the formulation of Lemma \ref{lemmacomposition} about the composition of advice: The advice space for the inner function in a composition has to be equipped with an injective representation, i.e.\ be strongly zero-dimensional and metrizable (\cite[Example 7.3.14]{engelking}) in order to obtain sensible results about the advice needed to compute the composition. From the spaces mentioned explicitly in the present paper, this criterion rules out Sierpi{\'n}ski space and $\mathbb{R}$. For the latter, however, a work-around exists: Using results such as Theorem \ref{theorandomintcant} or \cite[Corollary 4.9]{paulybrattka}, various kinds of $\mathbb{R}$-advice are equivalent to $Z$-advice for certain subspaces $Z \subseteq \Baire$, which in turn allows us to apply Lemma \ref{lemmacomposition}.

We conclude this section by giving three very simple functions that form distinct equivalence classes for direct reducibility, but collapse into one regarding Weihrauch reducibility. These are $f_1 : \mathbb{Q} \to \{0, 1\}$ with $f_1(0) = 1$ and $f_1(x) = 0$ otherwise; $f_2 : \mathbb{R} \to \mathbb{R}$ with $f_2(0) = 1$ and $f_2(x) = 0$ otherwise; $f_3 : \{(0, 0\} \cup \{(x, \sin(x^{-1})) \in \mathbb{R}^2 \mid 0 < x \leq 1\} \to \{0, 1\}$ with $f_3(0, 0) = 1$ and $f_3(x, y) = 0$ otherwise. The separation regarding direct reducibility stems from the fact that the domains are totally disconnected, path-connected and connected, but not path-connected, details can be found in \cite{paulymaster}. This separation, though, is lacking any algorithmic content: All three functions are semi-decidable, i.e.\ become computable, if the discrete space $\{0, 1\}$ is replaced by the Sierpi{\'n}ski space with $\{0\}$ as the non-trivial open set.

\section{Random Advice}
\label{secrandom}
Among the possible types of advice listed in Section \ref{secdefinitions}, random advice stands out as giving rise to a non-trivial concept that has not been studied in the literature so far. In this section, we will investigate its properties further, and draw some connections to \emph{Weak Weak K\"{o}nigs Lemma} from reverse mathematics.

As a starting point, our usual advice spaces shall be equipped with measures. For Cantor space $\Cantor$, we use the fair coin (uniform) measure, on $\mathbb{R}$ the usual Lebesgue measure is suitable. Finite spaces shall be equipped with a uniform measure. On $\mathbb{N}$ and $\Baire$ the situation is more complicated. However, the only relevant property of the measure is the set of null-sets. The measures induced by $p(n) = 2^{-n-1}$ on $\mathbb{N}$ and $\Baire$ give rise to a rather natural notion of null-sets.

Choosing Cantor space as a foundation, Baire space, $\mathbb{N}$ and the discrete finite spaces have standard representations $\delta$, such that the respective measure of a subset of the target space is identical to the fair coin measure of the set of $\delta$-names. This not only justifies the choice of measure on these spaces, this gives us:

\begin{proposition}
If $f$ is computable with random $Z$-advice, where $Z$ is $\mathbb{N}$, $\Baire$ or a finite discrete space, then $f$ is computable with random $\Cantor$-advice.
\end{proposition}

As all non-empty subsets of $\mathbb{N}$ or finite sets have positive measure, in these cases computability with advice is equivalent to computability with random advice, increasing the applicability of the proposition above.

Establishing the relationship between random $\Cantor$-advice and random $\mathbb{R}$-advice seems more complicated. In \cite{archibald} it was demonstrated that the measure induced on $\mathbb{R}$ by the uniform measure via the signed-digit representation is incomparable to the Lebesgue measure, that is there are null-sets for one measure with positive measure for the other in both directions. The natural representations in order to induce the Lebesgue measure are the conventional digit representations \cite{hertling2}, which are not equivalent to the standard representation. Using the unit-interval $\mathbb{I}$ rather than all of $\mathbb{R}$, we nevertheless arrive at the following conclusion:

\begin{theorem}
\label{theorandomintcant}
A function $f : \subseteq X \mto Y$ is computable with random $\Cantor$-advice, iff it is computable with random $\mathbb{I}$-advice.
\begin{proof}
Assume that $f$ is computable with random $\mathbb{I}$-advice, witnessed by $F$ and the sets $A_x \subseteq \mathbb{I}$ for $x \in \dom(f\delta_X)$. Let $\rho_2 : \Cantor \to \mathbb{I}$ denote the binary representation, considered as a function between represented spaces. It is computable, as $\rho_2$-names can be translated into $\rho$-names, and it preserves the measure. In particular, the sets $\rho_2^{-1}(A_x)$ all have positive measure. Now let $R$ be a computable realizer of $\rho_2$. Then $F' := F \circ (\id, R)$ and the sets $\rho_2^{-1}(A_x)$ witness computability of $f$ with random $\Cantor$-advice.

Now assume that $f$ is computable with random $\Cantor$-advice, witnessed by $F$ and the sets $A_x \subseteq \mathbb{I}$ for $x \in \dom(f\delta_X)$. Let $Q \subseteq \mathbb{I}$ be a Smith-Volterra-Cantor (or fat Cantor) set, that is a subset of $\mathbb{I}$ with positive measure homeomorphic to $\Cantor$. The usual construction directly yields a computable homeomorphism $\varphi :\subseteq \mathbb{I} \to \Cantor$ with $\dom(\varphi) = Q$ and $\lambda(\varphi^{-1}(A)) = \lambda(Q)\mu(A)$, where $\lambda$ is the Lebesgue measure on $\mathbb{I}$, and $\mu$ the uniform measure on $\Cantor$. Let $P$ be a computable realizer of $\varphi$. Then $F' := F \circ (\id, P)$ and the sets $\varphi^{-1}(A_x)$ witness computability of $f$ with random $\Cantor$-advice.
\end{proof}
\end{theorem}

\begin{corollary}
\label{corollaryrandomintcant}
$PC_{\Cantor} \equiv_W PC_{\mathbb{I}}$.
\begin{proof}
In addition to the proof of Theorem \ref{theorandomintcant}, note that the functions $\rho_2^{-1} : \mathcal{C}(\mathbb{I}) \to \mathcal{C}(\Cantor)$ and $\varphi^{-1} : \mathcal{C}(\Cantor) \to \mathcal{C}(\mathbb{I})$ are computable.
\end{proof}
\end{corollary}

In the next step, random advice shall be separated from the other advice concepts.
\begin{theorem}
\label{wklnotguessable}
$C_{\Cantor}$ is not computable with random $\Cantor$-advice.
\begin{proof}
Define $\Sep : \subseteq \Cantor \times \Cantor \to \Cantor$  via $(x, y) \in \dom(\Sep)$, if $x(i) = 0 \vee y(i) = 0$ holds for all $i \in \mathbb{N}$, and $z \in \Sep(x, y)$, if $x(i) = 1$ implies $z(i) = 1$ and $y(i) = 1$ implies $z(i) = 0$ for all $i \in \mathbb{N}$. Intuitively, $\Sep$ takes enumerations of two disjoint sets $A, B \subseteq \mathbb{N}$ and produces a separating set. Due to \cite{gherardi} we know $C_{\Cantor} \equiv_W \Sep$, so it is sufficient to show that $\Sep$ is not computable with random advice.

Now assume that a computable function $g$ witnesses the computability with random advice of $\Sep$. We pick $x, y \in \Cantor$ such that $\{i \in \mathbb{N} \mid x(i) = 1\}$ and $\{i \in \mathbb{N} \mid y(i) = 1\}$ are recursively enumerable sets that are not recursively separable. As $x$ and $y$ are computable, and so is the function $g$, for each $w \in \Cantor$ we have $g(\langle x, y\rangle, w) \leq_T w$, where $\leq_T$ denotes Turing reducibility. In particular, the set of suitable advice for input $x, y$ is a subset of: $$A_{\langle x, y\rangle} := \{w \in \Cantor \mid \exists z \in \Sep(x, y) \ z \leq_T w\}$$

However, \cite[Theorem 5.3]{jockusch} states $\lambda(A_{\langle x, y\rangle}) = 0$, refuting the assumption that $\Sep$ might be computable with random advice.
\end{proof}
\end{theorem}

As computability with $\mathbb{N}$-advice is trivially equivalent to computability with random $\mathbb{N}$-advice, it implies in turn computability with random $\Cantor$-advice. In order to separate the two notions, we will consider the combination with effective advice in the next step. Here, we can distinguish between effective random $\Cantor$-advice, effective random $\mathbb{R}$-advice and effective random $\Baire$-advice.

As a special case of Theorem \ref{effectiveadvicechoice}, exactly those functions are computable with effective random $\Cantor$-advice that are Weihrauch reducible to $PC_{\Cantor}$, the restriction of closed choice in Cantor space to sets with positive measure. Thus, the following proposition yields a separation of effective random $\Cantor$-advice and $\mathbb{N}$-advice:

\begin{proposition}
\label{wwklnotdiscrete}
$PC_{\Cantor}$ is not computable with $\mathbb{N}$-advice.
\begin{proof}
Let $(U_n)_{n \in \mathbb{N}}$ be a universal Martin L\"of test on $\Cantor$, and consider the set $\Cantor \setminus U_1$: It is a computable element of $\mathcal{C}(\Cantor)$ and it has a measure of at least $\frac{1}{2}$. In particular, all points in this set are random, thus, non-computable. Hence, $PC_{\Cantor}$ is not non-uniformly computable. Proposition \ref{nonuniform} completes the proof.
\end{proof}
\end{proposition}

\begin{figure}[htbp]
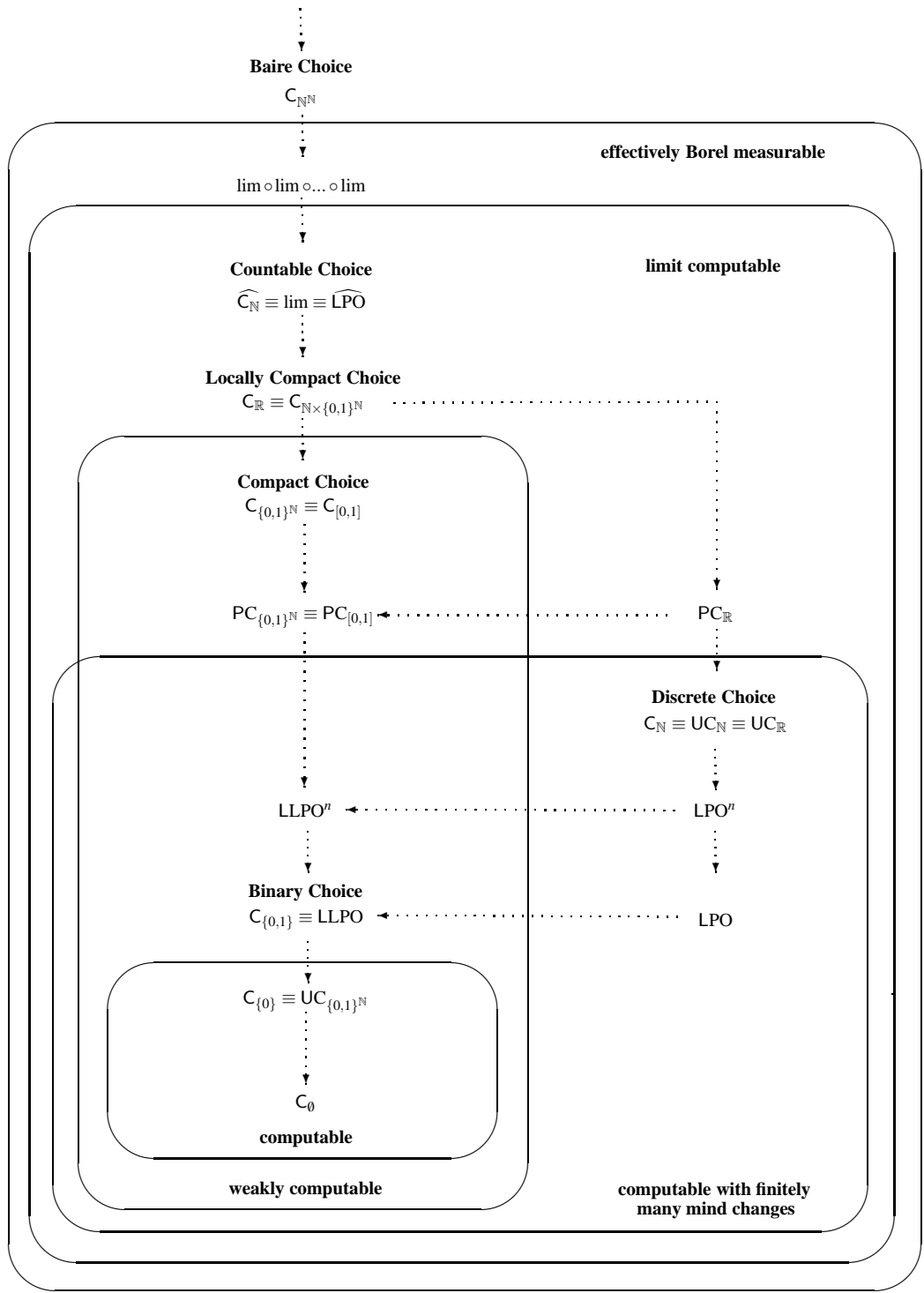

\begin{center}
\begin{scriptsize}
\input choice.pic
\end{scriptsize}
\caption{Choice principles in the Weihrauch lattice}
\label{fig:choice}
\end{center}
\end{figure}

The reducibility results obtained so far are already sufficient to describe the relative position of $PC_\mathbb{R}$ and $PC_{\Cantor}$ in the Weihrauch lattice compared to various other choice principles. $A |_W B$ denotes the incomparability of $A$ and $B$ regarding Weihrauch reducibility. The results are summarized in Figure \ref{fig:choice}, compare also \cite[Figure 1]{paulybrattka} and \cite[Figure 1]{brattka3}.
\begin{proposition}
$C_\mathbb{N} <_W PC_\mathbb{R}$, $C_{\Cantor} |_W PC_{\mathbb{R}}$, $PC_{\Cantor} |_W C_\mathbb{N}$, $PC_{\Cantor} <_W PC_\mathbb{R} <_W C_\mathbb{R}$.
\begin{proof}
$C_\mathbb{N}$ is reducible to $PC_\mathbb{R}$: The interval $[n, n + 0.5]$ is used as advice instead of $\{n\}$. $C_\mathbb{N}$ is incomparable with $C_{\Cantor}$, as shown in \cite{brattka3}. By transitivity, $PC_\mathbb{R}$ cannot be reducible to $C_{\Cantor}$.

As stated in Corollary \ref{corollaryrandomintcant}, $PC_{\Cantor}$ is equivalent to $PC_\mathbb{I}$, and the latter is by inclusion reducible to $PC_\mathbb{R}$. Using a homeomorphism\footnote{Note that this
    homeomorphism does not map closed subsets of $\mathbb{R}$
    to closed subsets of $[0,1]$ and hence it cannot be used
    to prove $PC_\mathbb{R} \equiv_W PC_\mathbb{I}$.} between the open interval $(0, 1)$ and $\mathbb{R}$, a consequence of Theorem \ref{theorandomintcant} is that $PC_\mathbb{R}$ is computable with random $\Cantor$-advice. By Theorem \ref{wklnotguessable} this is false for $C_{\Cantor}$, showing that $C_{\Cantor}$ cannot be reducible to $PC_\mathbb{R}$.

Finally, Corollary \ref{wwklnotdiscrete} implies $PC_{\Cantor} \nleq_W C_\mathbb{N}$. The remaining statements follow via transitivity.
\end{proof}
\end{proposition}

\def\C{{\rm\mathsf C}}
\def\UC{{\rm\mathsf UC}}
\def\PC{{\rm\mathsf PC}}
\def\LPO{{\rm\mathsf LPO}}
\def\LLPO{{\rm\mathsf LLPO}}
\def\IN{\mathbb{N}}
\def\IR{\mathbb{R}}
\def\Baire{\mathbb{N}^\mathbb{N}}
\def\Cantor{\{0,1\}^\mathbb{N}}

While $C_{\Cantor}$ basically is Weak K\"{o}nigs Lemma, $PC_{\Cantor}$ is called Weak Weak K\"{o}nigs Lemma ($WWKL_0$) in reverse mathematics (\cite{yu}, \cite{simpson3}). In the framework of reverse mathematics, several result from measure theory turn out to be equivalent to $WWKL_0$. As many of these are not existence theorems (e.g. additivity for some measure), they do not have counterparts as multi-valued partial functions, and cannot be treated in the framework of computability theory. Exceptions, however, are Vitali's Covering Theorem and a form of the Lebesgue monotone convergence theorem (\cite{yu2}).

The latter cannot be Weihrauch equivalent to $PC_{\Cantor}$, as every restriction of the Lebesgue convergence theorem is single-valued, and the Weihrauch degree of $PC_{\Cantor}$ does not contain any single-valued functions. Moreover, every single-valued function on computable metric spaces that is computable with effective $\Cantor$-advice is already computable, as shown in \cite{brattka2}.

Vitali's Covering Theorem turns out to be computable, if e.g. formulated for intervals. The proof of \cite[Lemma 5.3]{simpson3} has already algorithmic nature, $WWKL_0$ is only used to show that the algorithm actually works. As our framework allows the unrestricted use of classical mathematics to prove correctness of algorithms, Vitali's Covering Theorem is not Weihrauch equivalent to $PC_{\Cantor}$.

\section{Advice and Computational Complexity}
Expressing computational models such as Finitely Revising Type-2 Machines or Non-deterministic Type-2 Machines in terms of computation with certain advice opens a way to define computational complexity in these models:

\begin{definition}
\label{defcomplexity}
Let $f$ be computable with $(Z, \mathcal{A})$-advice. Define the complexity of $f$ with $(Z, \mathcal{A})$-advice as the minimal complexity of a function $F$ witnessing the computability of $f$ with $(Z, \mathcal{A})$-advice.
\end{definition}

Identical definitions are assumed for all kinds of additional restrictions. In order to fill Definition \ref{defcomplexity} with life, we have to specify what computational complexity is meant to be, and to show that such a \emph{minimal complexity} is a somehow well-defined concept.

Rather than attempting to fulfill this task in general, we refer to \cite{weihrauchf} for an introduction to complexity theory regarding computable metric spaces; and continue to study two specific cases, where interesting results can be obtained with a minimum of technical involvement.

If we restrict our attention to total single-valued functions $f : \Cantor \to \Cantor$, the time complexity of a function $f$ computed by a Type-2 Machine $M$ is defined as: $$\tau_M^f(k) := \max_{p \in \Cantor} \{n \in \mathbb{N} \mid M \textnormal{ needs } n \textnormal{ steps to produce the first } k \textnormal{ bits of } f(p)\}$$
As Cantor space is compact and $f$ total, this maximum always exists. As in the classical case, the complexity is a function $\tau: \mathbb{N} \to \mathbb{N}$, which allows us to define complexity classes such as $\mathbf{P}$ via: $$\mathbf{FP} := \{f : \Cantor \to \Cantor \mid \exists M \ \exists c, d \in \mathbb{N} \ \forall k \in \mathbb{N} \ \tau_M^f(k) \leq ck^d\}$$

As we will demonstrate now, we can also define $\mathbf{NP}$, using the equivalence of non-deterministic Type-2 machines (with alphabet $\{0, 1\}$) to computation with effective $\Cantor$-advice. We have already seen that, in contrast to the classical case, a non-deterministic Type-2 Machine is more powerful than a deterministic one; seemingly rendering any comparison of complexity classes futile.

However, the situation here is somewhat similar to the study of non-determinism in the BSS-model (\cite{blum2}): As long as functions are considered, non-deterministic BSS-machines are far more powerful than deterministic BSS-machines. A simple example is the function $x \mapsto \sqrt{x}$, more complicated ones are exhibited in \cite{paulybss}. However, when only decision problems are concerned, problems decidable in non-deterministic polynomial time are also decidable in deterministic time. Whether or not polynomial deterministic time is sufficient is an ongoing research topic.

In our case, the extra power of non-determinism over some finite alphabet can only be used for multi-valued functions: Any single-valued function computable with effective $\Cantor$-advice is already computable as a consequence of \cite[Corollary 8.8]{brattka2}. Thus, we proceed to define $\mathbf{FNP}$, exploiting the equivalence of $\Cantor$ and $\Cantor \times \Cantor$:

$$\mathbf{FNP} := \left \{f : \Cantor \to \Cantor \mid \begin{array}{l} \exists g \in \mathbf{FP},\ \exists \textnormal{computable } A : \Cantor \to \mathcal{C}(\Cantor) \\ \forall p \in \Cantor \ \forall r \in A(p) \ f(p) = g(\langle p, r\rangle)\end{array} \right\}$$

\begin{problem}
Is $\mathbf{FP} = \mathbf{FNP}$?
\end{problem}

Unlike the corresponding questions in the classical Turing machine model or the BSS-machines, a negative answer is readily obtained by the following lemma:

\begin{lemma}
$\mathbf{FNP} = \{f : \Cantor \to \Cantor \mid f \textnormal{ is computable}\}$
\begin{proof}
The inclusion $\subseteq$ was already stated, we will proceed to show the $\supseteq$-direction. Thus, we let $f$ be any computable function $f : \Cantor \to \Cantor$. Then $F: \Cantor \to \mathcal{C}(\Cantor)$ defined via $F(x) = \{f(x)\}$ is also computable. Therefore, $\pi_2 : \Cantor \times \Cantor \to \Cantor$ defined via $\pi_2(x, y) = y$ is a witness for the computability with effective $\Cantor$-advice of $f$. It is straight-forward to realize $\pi_2 \in \mathbf{FP}$.
\end{proof}
\end{lemma}

\begin{corollary}
$\mathbf{FP} \neq \mathbf{FNP}$
\end{corollary}

In order to prevent this rather disappointingly easy result, it would be necessary to take the complexity needed to compute the sets $A_x$ from $x$ into account. However, this would only apply to advice spaces such as $\mathbb{C}$, $\mathbb{R}$, $\mathbb{N}$, but exclude finite advice spaces.

For finite spaces, however, Definition \ref{defcomplexity} appears to exhibit nicer properties. In particular, access to finite advice does not influence the complexity of an already computable function $f : \Cantor \to \Cantor$. This stems from the fact that finite advice cannot be used to circumvent the computation of $f(x)$, as one of the values $g(x, 1)$, $g(x, 2)$, \ldots has to be identical to it.

With this, we have added a powerful argument to the ones listed in \cite[Remark 7]{ziegler5}: Computation with discrete advice, intended as a theory of non-uniform complexity, is compatible with uniform complexity.

\section{Additional Examples}
In the introduction we claimed that the interplay of Weihrauch reducibility and computability with advice can make proofs easier in both directions. To fortify our claim, we will give a few exemplary proofs, mainly of results presented originally in \cite{ziegler5}.

First, we consider the problem $LLPO : \subseteq \Cantor \times \Cantor \mto \{0, 1\}$ defined via $\dom(LLPO) = \{(x_0, x_1) \mid |\{(i, n) \in \{0, 1\} \times \mathbb{N} \mid x_i(n) = 1\}| \leq 1\}$ and $i \in LLPO(x_0, x_1)$, if $x_i = 0^\mathbb{N}$. The $n$-fold product of $LLPO$ with itself shall be denoted as $LLPO^n$.

\begin{proposition}
$LLPO^n$ is not computable with $\{1, \ldots, n-1\}$-advice\footnote{Already $\{1, \ldots, n\}$-advice is sufficient, as is the case for $LPO^n$. For $LPO^n$, the number of inputs equal to $0^\mathbb{N}$ is sufficient to determine which inputs are $0^\mathbb{N}$.}
\begin{proof}
As all involved spaces admit injective representations, if $LLPO^n$ were computable with $\{1, \ldots, n-1\}$-advice, then $LLPO^n$ would also admit a choice function that is computable with $\{1, \ldots, n-1\}$-advice. Now $LPO^n$ is reducible to each choice function of $LLPO^n$, thus, $LPO^n$ would be computable with $\{1, \ldots, n-1\}$-advice. Finally, $LPO^n$ is equivalent to $LPO_n$, and the latter was shown not to be computable with $\{1, \ldots, n-1\}$-advice in \cite{weihrauchc}.
\end{proof}
\end{proposition}

While $LLPO^n$ is of rather technical interest only, we can use this result to reprove a result from \cite{ziegler5}. For that, define $$\textsc{SEigen}_n : \subseteq \mathbb{R}^{n \times n} \mto \mathbb{R}$$ where $A \in \dom(\textsc{SEigen}_n)$, iff $A$ is symmetric, and $v \in \textsc{SEigen}_n(A)$, iff $v$ is an eigenvector of $A$ with $||v|| = 1$. \cite[Theorem 42]{ziegler5} states that $\textsc{SEigen}_n$ is not computable with $\{1, \ldots, \lfloor \log n \rfloor - 1\}$-advice. This\footnote{Given that our notion corresponds to weak advice in the terms of \cite{ziegler5}, our negative result is actually even stronger.} also follows from the following:

\begin{proposition}
\label{propeigen}
$LLPO^n \leq_W \textsc{SEigen}_{2^n}$.
\begin{proof}
First, we show $LLPO \leq_W \textsc{SEigen}_2$. For that, we use the matrices $$A = \left ( \begin{array}{cc} 1 & 0 \\ 0 & 2\end{array} \right ) \ \ \ \ B = \left ( \begin{array}{cc} 0 & 1 \\ 1 & 0 \end{array} \right )$$ We can consider the input to $LLPO$ to be real numbers rather than elements in Cantor space, thus, if we want to solve $LLPO(x, y)$, we apply $\textsc{SEigen}_2$ to $xA + yB$. If the result is neither $\pm \left ( \begin{array}{c}1 \\ 0 \end{array} \right )$ nor $\pm \left ( \begin{array}{c}0 \\ 1 \end{array} \right )$, then $y$ was not $0$, thus $0$ is a correct answer for $LLPO(x, y)$. If the result of $\textsc{SEigen}_2$ is neither $\pm 2^{-0.5} \left ( \begin{array}{c}1 \\ 1 \end{array} \right )$ nor $\pm 2^{-0.5} \left ( \begin{array}{c}1 \\ - 1 \end{array} \right )$, then $x$ was not $0$, thus $1$ is a correct answer for $LLPO(x, y)$. This completes the first reduction.

For the second step, simply observe $\textsc{SEigen}_{n} \times \textsc{SEigen}_{m} \leq_W \textsc{SEigen}_{nm}$. When searching for some eigenvector of the matrices $A$ and $B$, find an eigenvector $u \bigotimes v$ of $A \bigotimes B$ instead, and return the pair $(u, v)$. Here, $\bigotimes$ denotes the tensor product (or the Kronecker product), see e.g. \cite[Chapter 13]{laub}, in particular Theorem 13.12.
\end{proof}
\end{proposition}

As \cite[Theorem 42]{ziegler5} further states that $\textsc{SEigen}_n$ is computable with $\{1, \ldots, \lfloor \log n \rfloor\}$-advice, Proposition \ref{propeigen} is optimal.

Another example can be obtained via reduction from $MLPO_n$. We will define $MLPO_n : \subseteq \mathbb{R}^n \mto \{1, \ldots, n\}$ via $i \in MLPO_n(x_1, \ldots, x_n)$, if $x_i = 0$. With that, we have $MLPO_2 \equiv_W LLPO$. Using results from \cite{weihrauchc}, one can obtain that $MLPO_n$ is not computable with $\{1, \ldots, n - 1\}$-advice.

The problem to be considered is $\textnormal{LinEq}_{nm}$ defined in \cite[Theorem 35]{ziegler5}, which given some $n \times m$-matrix $A$ with $\operatorname{rank} A \leq \min \{n, m-1\}$, produces a vector $v \in \mathbb{R}^m \setminus \{0\}$ with $Av = 0$. The part of the result relevant for us is that $\textnormal{LinEq}_{nm}$ is not computable with $\{1, \ldots, \min \{n, m-1\} - 1\}$-advice. The most interesting case is $m = n+1$, and also covered by the following:

\begin{proposition}
$MLPO_{n+1} \leq_W \textnormal{LinEq}_{n,n+1}$ for $d = n = m-1$.
\begin{proof}
Let $(x_1, \ldots, x_{n+1})$ be the argument for $MLPO_{n+1}$. Construct the $n \times (n+1)$-matrix $A$ via $A_{ii} = x_i$ and $A_{i,i+1} = x_{i + 1}$ for $1 \leq i \leq n$, and $A_{kl} = 0$ else. Apply $\textnormal{LinEq}_{n,n+1}$ to $A$ to obtain a vector $v \neq 0 \in \mathbb{R}^{n+1}$. Search for an index $i$ of a non-zero component of $v$.

Assume $x_i \neq 0$. Because we have $x_{i - 1}v_{i - 1} + x_iv_i = 0$, also $x_{i - 1} \neq 0$ follows (provided $i > 1$), in the same way, we can conclude $x_{i + 1} \neq 0$ (provided $i \leq n$). By iterating this, we obtain $x_j \neq 0$ for all $1 \leq j \leq n+1$, a contradiction to the assumption that $(x_1, \ldots, x_{n+1})$ is in the domain of $MLPO_{n+1}$. Thus, $x_i = 0$ must be true, and $i$ is a valid output.
\end{proof}
\end{proposition}

\end{document}